\documentclass[pra,superscriptaddress,showpacs,onecolumn,floatfix]{revtex4}
\usepackage{amssymb}
\usepackage{amsfonts}
\usepackage{amsmath}
\usepackage{graphicx}
\begin{document}
\title{Cloning of spin coherent states}
\author{Rafa{\l} Demkowicz-Dobrza\'nski}
\affiliation{Center for Theoretical Physics, Polish Academy of Sciences, Aleja Lotnik{\'o}w 32/44, 02-668 Warszawa, Poland}
\affiliation{Institute of Theoretical Physics, Warsaw University, Warszawa, Poland}
\author{Marek Ku{\'s}}
\affiliation{Center for Theoretical Physics, Polish Academy of Sciences, Aleja Lotnik{\'o}w 32/44, 02-668 Warszawa, Poland}
\affiliation{Faculty of Mathematics and Sciences, Cardinal Stefan Wyszyñski University, Warszawa, Poland}
\author{Krzysztof W\'odkiewicz}
\affiliation{Institute of Theoretical Physics, Warsaw University, Warszawa, Poland}
\date{\today}

\begin{abstract}
We consider optimal cloning of the spin coherent states in Hilbert spaces of
different dimensionality $d$. We give explicit form of optimal cloning
transformation for spin coherent states in the three-dimensional space,
analytical results for the fidelity of the optimal cloning in $d=3$ and $d=4$
as well as numerical results for higher dimensions. In the low-dimensional case
we construct the corresponding completely positive maps and exhibit
their structure with the help of Jamio{\l}kowski isomorphism. This allows us
to formulate some conjectures about the form of optimal coherent cloning
CP maps in arbitrary dimension.
\end{abstract}
\maketitle

\section{Introduction}
The no-cloning theorem \cite{wooters,yuen,barnum} claims that a universal and
faithful cloning machine that would clone perfectly an arbitrary input quantum
state, is incompatible with quantum mechanics. It is possible, however, to find
imperfect cloning machines that would copy quantum states with some loss of
quality. The interesting question is what are the optimal cloning machines
allowed by quantum mechanics that would clone quantum states as good as
possible?

Recently, many papers have appeared describing different optimal universal
cloning machines designed to clone arbitrary states from a Hilbert space of
dimension $d$. Optimal schemes for cloning of pure states, giving the highest
possible fidelity, were found for $1 \to 2$ (two copies of the original input
state are produced) and more generally $N \to M$ ($M$ copies of $N$ identical
input states) universal cloning in arbitrary dimension
\cite{buzek,gisin,db:clon,rw:clon,hf:clon}.

One can consider also nonuniversal cloning machines designed specially for
cloning certain subsets of states from a given Hilbert space. In this approach
one attempts to optimize fidelity for cloning states from a subset without
taking care how well other states (outside the subset) are cloned. Usually one
also imposes a condition that all states from the chosen subset are cloned
equally well.

The case when one wants to find the best possible machine for cloning of two
given nonorthogonal states was discussed in \cite{nonort}. Cloning of coherent
states in an infinite dimensional space was considered in \cite{nc:coh,sb:coh},
where transformation for cloning of all coherent states with the same fidelity
$\mathcal{F}=2/3$ was proposed. A proof that such a cloning procedure gives
the highest possible fidelity for coherent states was later given in
\cite{nc:optimal}.

The proof, however, was
not a straightforward one and limited to the case of so called gaussian cloners. By a straightforward proof
we mean a standard optimization procedure, where one starts with the most general
unitary transformation on an input state, a blank state and an ancilla, adds
unitary constraints, and then tries to optimize fidelity for cloning of the
desired group of states. This seems to be very difficult to do in the case of
optimizing cloning of coherent states, as these are states taken from infinite
dimensional Hilbert space.

Instead of the optimization scheme, the authors of \cite{nc:optimal} make use
of limitations on so called \emph{joint measurement} \cite{joint}. The idea of
joint measurement is to perform some kind of a measurement on a quantum
system that would give us simultaneously some information about two noncommuting
observables (for example position and momentum). One way of performing joint
measurement can be realized with the help of quantum cloning \cite{jointclon}.
In order to measure simultaneously position and momentum of a particle, one can
first clone its quantum state and then perform a position measurement on one
of the clones and a momentum measurement on the other. As cloning must not be
a way to circumvent fundamental limitations on accuracy of a joint measurement,
authors arrive at the conclusion that the maximal attainable fidelity for cloning
of coherent states must satisfy: $\mathcal{F} \leq 2/3$.

In this paper we shall consider spin coherent states in a finite dimensional
space and we shall look for the optimal cloning transformation for them. In the
limit of $d \to \infty$ the coherent states we use tend to the harmonic
oscillator coherent states. Therefore for higher dimensions we shall be able to
compare our results on maximal possible fidelity with that based on joint
measurement arguments.

In section \ref{sec:cp} we discuss the same problem of cloning of spin coherent states
using different approach. 
We analyze completely positive (CP) maps corresponding to the
universal and the coherent cloning transformation in the case of $d=2$, $d=3$.
In doing so, we follow the general approach, proposed by D'Ariano and Presti
\cite{ga:nonuniversal} to the problem of nonuniversal covariant cloning. We
exhibit the structure of cloning CP maps using Jamio{\l}kowski isomorphism
\cite{aj:positive}. This section is intended to give a better insight into the form of the
optimal coherent cloning transformation in low dimensions, which we believe may be
helpful in finding an analytical solution for the problem of cloning of spin coherent states 
 in an arbitrary dimension $d$.

\section{Coherent states}

In the infinite dimensional space of the harmonic oscillator states we can
construct the creation and annihilation operators, $a^\dagger,a$, obeying the
boson commutation relation $[a,a^\dagger]=1$. The coherent states of such a
system (harmonic oscillator coherent states) are eigenvectors of the
annihilation operator $a|\alpha\rangle =\alpha|\alpha\rangle $ and can be
obtained as displacements of the ground state $|0\rangle $:
\begin{equation}
|\alpha\rangle =D(\alpha)|0\rangle , \quad  D(\alpha)=
\exp(\alpha a^\dagger-\alpha^*a).
\end{equation}
One of important properties of such coherent states is that they satisfy the
lower bound on product of the dispersions of the position and momentum
operators (or the quadrature operators) required by the Heisenberg uncertainty
principle.

The concept of coherent states is not restricted to the infinite dimensional
space. In a finite dimensional space one can introduce different kinds of
coherent states \cite{perolomov}. In this paper we shall concentrate on so
called spin coherent states ($SU(2)$ coherent states), which we define below
\cite{spincoh}.

Let us consider the Hilbert space of spin states with the total spin $j$, the
space in question has the dimension $d=2j+1$. By $|m\rangle $,
$m=-j,-j+1,\dots,j-1,j$ we denote the basis consisting of the eigenvectors of
$\hat{J}_z$ operator. Spin coherent states are defined as rotations of the
''ground'' state $|-j\rangle$ by unitary operators from the irreducible $SU(2)$
representation in the $2j+1$ dimensional space:
\begin{equation}
|\theta, \phi\rangle  = R_{\theta, \phi}|-j\rangle ,
\quad R_{\theta, \phi}=e^{-i\theta(\hat{J}_x \sin\phi -\hat{J}_y\cos\phi)}.
\end{equation}
The operator $R_{\theta, \phi}$ corresponds to a rotation by the angle $\theta$
around the axis $\vec{n}=[\sin\phi,-\cos\phi,0]$. For $j=1/2$ the dimension of
the space is $d=2$ (qubit). In this case spin coherent states are actually all
the pure states in the space (every pure state can be described by a direction
on the Bloch sphere). In higher dimensions, however, spin coherent states
constitute only a subset in the set of all states of a given Hilbert space.

Their similarity to the harmonic oscillator coherent states lies in the fact
that they are constructed as rotations of the ''ground'' state just like the
harmonic oscillator coherent states were constructed as displacements of the
ground state. For us the most important feature of spin coherent states is that
they approach harmonic oscillator coherent states when dimension of the space
tends to infinity \cite{spincoh}.

One can decompose a spin coherent state $|\theta,\phi\rangle $ in the
$\hat{J}_z$-eigenvectors basis:
\begin{equation}
\label{eq:decomposition}
\langle m|\theta,\phi\rangle = \left(
\begin{array}{c}
2j\\
j+m
\end{array}
\right)^{1/2}\sin^{j+m}(\theta/2)\cos^{j-m}(\theta/2)e^{-i(j+m)\phi}.
\end{equation}

\section{Optimal cloning of spin coherent states}

The most general $1 \to 2$ cloning transformation which clones pure states from
a $d$-dimensional Hilbert space $\mathcal{H}_1$, is a unitary transformation
$U$ acting on Hilbert space $\mathcal{H}_1 \otimes \mathcal{H}_2 \otimes
\mathcal{H}_a$. The $d$-dimensional spaces $\mathcal{H}_1$ and ${\mathcal
H}_2$ correspond to the input system and a blank one which after cloning
will carry cloned states, while $\mathcal{H}_a$ is the Hilbert space of
ancilla states.

We shall denote a basis in the space $\mathcal{H}_1$ (or
$\mathcal{H}_2$) by $|n\rangle $, $n=0,\dots,d-1$, where $|0\rangle $
corresponds to the ground state $|-j\rangle $. Coherent states in this space
are thus rotations of the state $|0\rangle $. We shall refer to the states
$|n\rangle $ as the number states.

The cloning machine will act on the state $|\psi\rangle _1|0\rangle
_2|A_0\rangle $, where $|\psi\rangle _1 \in \mathcal{H}_1$ is a state to be
copied, while the blank system and the ancilla are always prepared in the same
initial states denoted here by $|0\rangle _2|A_0\rangle $. The final state will
be $U|\psi\rangle _1|0\rangle _2|A_0\rangle$. After tracing out the output density
matrix with respect to the ancilla states and the states of one of the clones,
one obtains reduced density matrix for the remaining clone. In this paper only
symmetric cloning is considered, so the two reduced density matrices for the two
clones are the same and are denoted by $\rho^{\textrm{out}}$.

Looking for the best possible machine for cloning of spin coherent states from
space $\mathcal{H}_1$ means maximizing the average fidelity for cloning of
these states:
\begin{equation}
\label{eq:fidelity}
\mathcal{F}=\int d\Omega \langle \theta,\phi|\rho^{\textrm{out}}_{\theta,\phi}|\theta,
\phi\rangle,
\end{equation}
where $d\Omega=d\phi\, d\theta \sin\theta/ 4\pi$, and the
$\rho_{\theta,\phi}^{\textrm{out}}$ is the density matrix of the clones after cloning
the input spin coherent state $|\theta, \phi \rangle$.

In order to find the maximal attainable fidelity for cloning of spin coherent
states and corresponding cloning transformation we shall generalize the method
used by Gisin and Massar in \cite{gisin}.

Firstly we assume that the final state is symmetric with respect to the
exchange of the two clones. Secondly we assume that our cloning machine clones
all coherent states with the same fidelity. We may make these assumptions,
because they do not lower possible attainable fidelity, as was explained in
\cite{gisin}.

In $\mathcal{H}_1 \otimes \mathcal{H}_2$ space there are $S=d(d-1)/2$
symmetric states. We shall denote them by $|s\rangle _{12}$, $s=0 \dots S-1$.
The most general cloning transformation can thus be described as:
\begin{equation}
U|n\rangle _1|0\rangle _2|A_0\rangle =|s\rangle _{12}|R_{ns}\rangle ,
\end{equation}
where $|R_{ns}\rangle $ are unnormalized states of the ancilla system and
the summation convention is used. Unitarity of transformation imposes a set of
constraints:
\begin{equation}
\label{eq:constraints}
\langle R_{n's}|R_{ns}\rangle =\delta_{n'n}.
\end{equation}
Every coherent state can be decomposed in the basis of number states
$|\theta,\phi\rangle =O_n|n\rangle $. According to
Eq.~(\ref{eq:decomposition}):
\begin{equation}
O_n = \langle n| \theta, \phi\rangle =
\left(
\begin{array}{c}
d-1\\
n
\end{array}
\right)^{1/2}\sin^{n}(\theta/2)\cos^{d-n-1}(\theta/2)e^{-in\phi},
\end{equation}
where $d=2j+1$, $n=j+m$. Action of the cloning transformation $U$ on such a
coherent state gives:
\begin{equation}
U|\theta,\phi\rangle _1|0\rangle _2|0\rangle _a = O_n|s\rangle _{12}|R_{ns}
\rangle.
\end{equation}
Calculating the fidelity from equation (\ref{eq:fidelity}) one arrives at the
formula:
\begin{equation}
\begin{array}{rcl}
\mathcal{F} & = &\langle R_{n's'}|R_{ns}\rangle \langle k|\langle l|s\rangle
\langle s'|k'\rangle |l\rangle\int d\Omega \ O_{n'}^*O_{k'}O_{k}^*O_{n}=\\
\\
&=&\langle R_{n's'}|R_{ns}\rangle A_{n's'ns}.
\end{array}
\end{equation}
As explained in \cite{gisin}, due to the rotational symmetry of the cloning
machine, it is enough to consider only one constraint from the set
(\ref{eq:constraints}), namely $\langle R_{ns}|R_{ns}\rangle =d$. Incorporating
this constraint by means of the Lagrange multiplier $\lambda$ we have to
extremize the following expression:
\begin{equation}
\mathcal{F}=\langle R_{n's'}|R_{ns}\rangle A_{n's'ns}-
\lambda(\langle R_{n's'}|R_{ns}\rangle \delta_{s's}\delta_{n'n}-d).
\end{equation}
Denoting by $\{|a\rangle\}$ an orthonormal basis in the ancilla system, we can
insert the identity operator $\mathbb{I} = \sum_a|a\rangle \langle a|$ between
$\langle R_{n's'}|$ and $|R_{ns}\rangle $. Then, varying over $\langle
R_{n'a'}|a\rangle $, we obtain an eigenvalue problem:
\begin{equation}
\label{eq:eigen}
(A_{n's'ns}-\lambda\delta_{s's}\delta_{n'n})\langle a|R_{ns}\rangle =0.
\end{equation}
Multiplying on the left by $\langle R_{n's'}|a\rangle $ and summing over
$n',s'$, and $a$ one concludes that the fidelity is $\mathcal{F}=d \lambda$. 
Thus looking for the best fidelity is equivalent to looking
for the biggest eigenvalue of (\ref{eq:eigen}). From the eigenvectors
corresponding to this eigenvalue one can infer the form of the unitary
transformation which attains this fidelity.

\begin{figure}[htb]
\begin{center}
\includegraphics{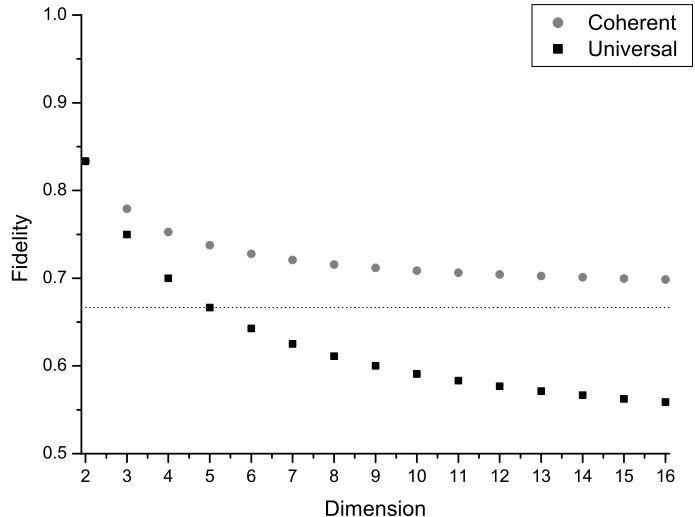}
\caption{Comparison between fidelity of coherent and universal
cloning machines.}
\label{fi:fidelities}
\end{center}
\end{figure}

In the case of the universal cloning machine in $d$ dimensions it was proven
\cite{rw:clon} that the optimal fidelity for $1 \to 2$ universal cloning reads:
\begin{equation}
\mathcal{F}^{\textrm{universal}}_d= \frac{d+3}{2d+2}.
\end{equation}
As we attempt to clone only coherent states from a given space, fidelity of
such cloning should be higher $\mathcal{F}^{\textrm{coherent}}_d \geq {\mathcal
F}^{\textrm{universal}}_d$ (equality holds only for $d=2$). We obtained analytical
solutions for the maximal fidelity of cloning of spin coherent states for $d=3$
and $d=4$:
\begin{equation}
\label{eq:exact}
\mathcal{F}^{\textrm{coherent}}_3=\frac{11+\sqrt{21}}{20},
\quad \mathcal{F}^{\textrm{coherent}}_4=\frac{79+\sqrt{697}}{140}.
\end{equation}
We have also calculated the numerical values of fidelities for higher
dimensions $d \leq 16$ (The fact that we stopped our calculations
at $d=16$, is merely because of the rapid growth of computation time with the increase of $d$.
(This is because the number of elements of matrix $A$, we had to find the biggest eigenvalue of, was increasing as $d^6$). 
 Comparison between results for the maximal fidelities
of the coherent and the universal cloning is shown in Figure
\ref{fi:fidelities}.

The dotted line marks the value of the fidelity equal to 2/3. Fidelity of
universal cloning machines tends to $1/2$ as the dimension tends to $\infty$,
and falls below $2/3$ as soon as for $d=6$. Our numerical results show that
with the increase of the dimension fidelity of the coherent cloning falls
significantly slower than that of the universal cloning. In our calculations it
has not fallen below $2/3$. For the highest dimension we calculated, $d=16$, we
obtained fidelity $\mathcal{F}^{\textrm{coherent}}_{16}=0.699$. These numerical results
are useful in giving limitations for the maximal possible fidelity for cloning
of harmonic oscillator coherent states, as for higher dimension spin coherent
states tend to harmonic oscillator ones. In our case we can infer that fidelity
for cloning of the harmonic oscillator coherent states satisfies ${\mathcal
F}^{\textrm{coherent}}_{\infty} \leq 0.699$. Calculation of fidelities for higher
dimension gives a better bound on this fidelity. From these results we
 cannot conclude, however, whether the fidelity $\mathcal{F}^{\textrm{coherent}}_{\infty}$ will 
actually reach the value $2/3$ or will it saturate somewhere above.

We have recently learned that using different approach and with the
help of some numerics, Navez and Cerf have found that the optimal 
fidelity for cloning of harmonic oscillator coherent states is $0.6825$ \cite{Navez},
which indeed falls into the range $[2/3, 0.699]$. 

For the sake of curosity we can try to fit to our numerical results, 
a rational function (with three free parameters $\alpha, \beta, \gamma$) of the form $\mathcal{F}^{\textrm{fit}}_d= \frac{\alpha d + \beta}{d + \gamma}$
(we know, however, that the exact formula for the optimal fidelity of cloning of spin coherent
states is not a rational function, as we have found analytical solutions for $d=3$ and $d=4$
(\ref{eq:exact}))). This fitting gives the asymptotic behaviour $\mathcal{F}^{\textrm{fit}}_{\infty} = 0.6812$,
which may be seen as an independent indication that optimal fidelity for cloning
of harmonic oscillator coherent states is indeed above $2/3$.

The lowest dimension for which universal cloning and coherent cloning differ
from each other is $d=3$. It is interesting to compare the explicit form of the
universal and the coherent cloning transformations in this case. Both
transformations require that the ancilla system is at least three dimensional.
We denote the three orthogonal ancilla states as $|A_0\rangle , |A_1\rangle ,
|A_2\rangle $. Optimal universal cloning transformation reads:
\begin{equation}
\left\{
\begin{array}{rcl}
|0\rangle |0\rangle |A_0\rangle  &\to&\frac{1}{\sqrt{2}}|0,0\rangle |A_0\rangle
+\frac{1}{2}|0,1\rangle |A_1\rangle +\frac{1}{2}|0,2\rangle |A_2\rangle \\
\\
|1\rangle |0\rangle |A_0\rangle  &\to&\frac{1}{\sqrt{2}}|1,1\rangle |A_1\rangle
+\frac{1}{2}|0,1\rangle |A_0\rangle +\frac{1}{2}|1,2\rangle |A_2\rangle \\
\\
|2\rangle |0\rangle |A_0\rangle  &\to&\frac{1}{\sqrt{2}}|2,2\rangle |A_2\rangle
+\frac{1}{2}|0,2\rangle |A_0\rangle +\frac{1}{2}|1,2\rangle |A_1\rangle ,\\
\end{array}\right.
\end{equation}
where $|i,i \rangle$ denotes $|i\rangle|i\rangle$ state while by $|i,j\rangle $ (for $i \neq j$)we mean the symmetrized state
$1/\sqrt{2}(|i\rangle |j\rangle +|j\rangle,|i\rangle)$. The optimal coherent cloning
transformation in this case is given by:
\begin{equation}
\left\{
\begin{array}{rcl}
|0\rangle |0\rangle |A_0\rangle  &\to&\alpha|0,0\rangle |A_0\rangle
+\frac{1}{\sqrt{2}}\alpha|0,1\rangle |A_1\rangle
+(\beta|1,1\rangle +\delta|0,2\rangle )|A_2\rangle \\
\\
|1\rangle |0\rangle |A_0\rangle  &\to&(\gamma|1,1\rangle +\sqrt{2}\beta|0,2\rangle
)|A_1\rangle +\frac{1}{\sqrt{2}}\alpha|0,1\rangle |A_0\rangle
+\frac{1}{\sqrt{2}}\alpha|1,2\rangle |A_2\rangle \\
\\
|2\rangle |0\rangle |A_0\rangle  &\to&\alpha|2,2\rangle |A_2\rangle
+\frac{1}{\sqrt{2}}\alpha|1,2\rangle |A_1\rangle
+(\beta|1,1\rangle +\delta|0,2\rangle )|A_0\rangle ,\\
\end{array}\right.
\end{equation}
with the coefficients:
\begin{equation}
\begin{array}{rcccl}
\alpha&=&\sqrt{\frac{1}{210}(63+13\sqrt{21})} &\approx& 0.764 \\
\\
\beta&=&\sqrt{\frac{1}{5}-\frac{4}{5\sqrt{21}})} &\approx& 0.159\\
\\
\gamma&=&\sqrt{\frac{1}{70}(21+\sqrt{21})} &\approx& 0.604\\
\\
\delta&=&\sqrt{\frac{7}{20}-\frac{23}{20\sqrt{21}}} &\approx& 0.315.
\end{array}
\end{equation}
Note that while in the case of the universal cloning none of basis states was
distinguished (all were cloned in the similar manner), in the case of the
coherent cloning the state $|1\rangle $ is cloned differently than two other
basis states. The reason for this is that the state $|1\rangle $ is not a spin
coherent state while states $|0\rangle $, $|2\rangle $ are.

We can compare the reduced density matrices for the two clones in the case of
the universal and the coherent cloning. With the state $|0\rangle $ at the
input reduced density matrices after cloning with the universal and the
coherent cloning machines are respectively given by:
\begin{equation}
\begin{array}{rcl}
\rho^{\textrm{universal}}_{\textrm{out}}&=& 0.750|0\rangle \langle 0|+0.125|1\rangle \langle 1|
+0.125|2\rangle \langle 2|\\
\\
\rho^{\textrm{coherent}}_{\textrm{out}}&=& (\frac{5}{4}\alpha^2+\delta^2)|0\rangle \langle 0|
+(\frac{1}{4}\alpha^2+\beta^2)|1\rangle \langle 1|+\delta^2|2\rangle \langle 2| \approx\\
\\
&\approx& 0.779|0\rangle \langle 0|+0.171|1\rangle \langle 1|+0.049|2\rangle \langle 2|.
\end{array}
\end{equation}
After the universal cloning the resulting reduced density matrix (for cloning
in any dimension) is a mixture of the initial density matrix $|\psi\rangle
\langle \psi|$ and the identity matrix of the appropriate dimension. That is
why $\rho^{\textrm{universal}}$ has equal contributions from $|1\rangle \langle 1|$ and
$|2\rangle \langle 2|$ and a dominant term with $|0\rangle \langle 0|$.

In case of a coherent cloning the state $|0\rangle $ (a coherent state) is
cloned better than by the universal machine (cf. higher coefficient in front of
$|0\rangle \langle 0|$). Additionally, other basis states $|n\rangle \langle
n|$ ($n \neq 0$) contribute with different coefficients, the bigger is $n$ the
lower is the coefficient.

It is in accordance with what we know about cloning of the harmonic oscillator
coherent states. When one clones a vacuum state through the optimal gaussian cloning
machine for coherent states, reduced density matrix of clones describes then a
thermal state with mean number of photons equal to $1/2$. It is a mixture of
number states $|n\rangle \langle n|$ with the coefficients decreasing when $n$
increases.

\section{CP map and positive operator approach}
\label{sec:cp}
In this section we do not present any new cloning transformations, 
nor new results for optimal cloning fidelities.    
Instead, we analyze the structure of completely positive (CP) maps corresponding to the
universal and the coherent cloning transformations obtained in previous section.
This section is intended to give a better insight into the form of the
optimal coherent cloning transformation in low dimensions, which we believe may be
helpful in finding an analytical solution for the problem of cloning of spin coherent states 
in an arbitrary dimension $d$.

The essential elements in the cloning process are the input state and the two
output clones. Therefore instead of considering cloning process as a unitary
operation $U$ on the full space $\mathcal{H}_1 \otimes \mathcal{H}_2 \otimes
\mathcal{H}_a$, one can consider a corresponding complete positive (CP), trace
preserving map $\mathcal{E}: \mathcal{L}(\mathcal{H}_1) \to {\mathcal
L}(\mathcal{H}_1 \otimes \mathcal{H}_2)$, where $\mathcal{L}(\mathcal{H})$
denotes a space of linear bounded operators on the Hilbert space ${\mathcal
H}$. The relation between the CP map $\mathcal{E}$ and the unitary
transformation $U$ is the following:
\begin{equation}
\mathcal{E}(\rho) = \textrm{Tr}_a(U\rho \otimes |0\rangle\langle 0| \otimes |0
\rangle \langle 0| U^\dagger),
\end{equation}
where $\rho=|\psi\rangle \langle \psi| $ is the state to be copied and the
trace is calculated with respect to the ancilla states.

Let $\mathcal{H}$ and $\mathcal{K}$ be two Hilbert spaces. For any CP map
from $\mathcal{L}(\mathcal{H})$ to $\mathcal{L}(\mathcal{K})$, one can
define a positive operator $P_\mathcal{E} \in \mathcal{L}(\mathcal{K}
\otimes \mathcal{H})$:
\begin{equation}
P_\mathcal{E} = \mathcal{E}\otimes \mathcal{I}(|{\mathbb I} \rangle \langle
\mathbb{I}|),
\end{equation}
where $|\mathbb{I}\rangle = \sum\limits_{n} |n \rangle \otimes |n \rangle $ is
an unnormalized state in $\mathcal{H} \otimes \mathcal{H}$ (maximally
entangled state), $|n \rangle$ are base vectors in $\mathcal{H}$, and
$\mathcal{I}$ is the identity map on $\mathcal{L}(\mathcal{H})$. This
definition gives one to one correspondence between CP maps and positive
operators known as the Jamio{\l}kowski isomorphism \cite{aj:positive}. As we
consider trace preserving CP maps, we must impose an additional condition on
$P_\mathcal{E}$:
\begin{equation}
\label{eq:trace}
\textrm{Tr}_\mathcal{K}(P_\mathcal{E})= \mathbb{I} \in \mathcal{L}(H).
\end{equation}

Cloning of spin coherent states is a special case of a nonuniversal covariant
cloning i.e. a cloning scheme covariant under a proper subgroup (all the states
generated by this subgroup will be cloned in the same way) of the unitary group
in d dimensions $U(d)$. In the case of the cloning of spin coherent states this
subgroup is the $SU(2)$ group. General approach to the problem of optimizing
nonuniversal covariant cloning using positive operator $P_\mathcal{E}$ picture
was presented in \cite{ga:nonuniversal}. We shall now look at our previous
results for cloning of spin coherent states from this point of view.

We are interested in the covariant cloning with respect to $SU(2)$ group in d
dimensional space. In terms of CP map this means that if $R_d$ is an
irreducible representation of $SU(2)$ in $d$ dimensional space then the
condition
\begin{equation}
\mathcal{E}\left(R_d(g) \rho R_d(g)^\dagger\right) = R_d(g)^{\otimes 2}
\mathcal{E}(\rho) R_d(g)^{\dagger \otimes 2},
\end{equation}
must be satisfied for any $g \in SU(2)$ and $\rho \in \mathcal{L}(\mathcal{H}_1)$.

The covariance condition for corresponding positive operator $P_\mathcal{E}\in
\mathcal{L}(\mathcal{H}_{1} \otimes \mathcal{H}_2 \otimes \mathcal{H}_1)$
reads \cite{ga:nonuniversal}:
\begin{equation}
\label{eq:commutation}
[P_\mathcal{E}, R_d(g)^{ \otimes 2}\otimes R^*_d(g)]=0,
\end{equation}
where $R_d^*$ denotes the complex conjugate representation. One can decompose
the tensor product of representations into a direct sum of irreducible
representation:
\begin{equation}
\label{eq:tensor}
R_d^{\otimes 2} \otimes R_d^* = \oplus_i R_d^i.
\end{equation}
The full Hilbert space is decomposed into invariant subspaces $\mathcal{H}_{1}
\otimes \mathcal{H}_2 \otimes \mathcal{H}_1 =  \oplus_i \mathcal{M}^i$.
The representation $R^i$ acts in the subspace $\mathcal{M}^i$.

The commutation condition (\ref{eq:commutation}) imposes a certain block
 structure on the operator $P_\mathcal{E}$:
\begin{equation}
P_\mathcal{E} = \sum \limits_{ij} c_{ij} \mathbb{I}^i_j,
\end{equation}
where $c_{ij}=0$ if representations $R^i$ and $R^j$ are inequivalent, and
$\mathbb{I}^i_j$ denotes isomorphism between spaces $\mathcal{M}^i$ and
$\mathcal{M}^j$ (which exists if $R^i$ and $R^j$ are equivalent). In order to
have a positive operator $P_\mathcal{E}$ each block must be a positive matrix.

As we are interested in the symmetric cloning, in addition to the covariance
with respect to the $SU(2)$ group we have to impose covariance with respect to
permutation of the two output clones. This is equivalent to commutativity of
the operator $P_\mathcal{E}$ and $S_2 \otimes \mathbb{I}$, where $S_2$
denotes the permutation of two clones (the permutation in the space $\mathcal{H}_1
 \otimes \mathcal{H}_2$). As a consequence $c_{ij}$ coefficients are zero
even if representations $R^i$, $R^j$ are equivalent, but states from the
corresponding spaces $\mathcal{M}^i$, $\mathcal{M}^j$ have different symmetry
(i.e.\ they are symmetric or antisymmetric) with respect to exchange of clones.
From now on we shall not distinguish between spaces $\mathcal{H}_1$ and
$\mathcal{H}_2$ and both will be denoted by $\mathcal{H}$.

In order to find explicitly the decomposition (\ref{eq:tensor}) of the tensor
product of three $d$-dimensional representations of $SU(2)$ group one has to
follow the scheme of adding angular momenta of three particles. We shall first
add angular momenta of the particles 1 and 2 using Clebsch-Gordan coefficients
and then add the third particle in the same manner \cite{jn:rotations}. We
choose this order because we want the states from invariant subspaces to have
definite symmetry with respect to the exchange of particles 1 and 2 (in our
case they correspond to the two clones).

The situation here is a little bit different, however, from the case of adding
angular momenta of three particles, as the third representation in tensor
product (\ref{eq:tensor}) is a complex conjugate one. Fortunately in the case
of representations of $SU(2)$ a complex conjugate representation $R^*$ is
always equivalent to $R$ and the only thing we have to modify when performing
the decomposition is to make a substitution $|n\rangle \to (-1)^n|d-n-1 \rangle$
for the basis states in the third space \cite{jn:rotations}.

\subsection{d=2}
As an illustration we shall first consider spin coherent cloning in the case
$d=2$ (universal cloning of qubit). The decomposition (\ref{eq:tensor}) reads
now:
\begin{equation}
R_2 \otimes R_2 \otimes R_2^* = R_2 \oplus R_2 \oplus R_4.
\end{equation}
In the language of angular momentum this means that adding three particles
with spin $1/2$ we have two invariant spaces corresponding to spin $1/2$ and
one space corresponding to spin $3/2$. The Hilbert space can be decomposed into
three invariant subspaces $\mathcal{H}^{\otimes 2+1}=\oplus_{i=1}^3 \mathcal{M}^i$. 
This decomposition is shown in Table \ref{ta:d2}.

\begin{table}[htb]
\caption{\label{ta:d2}Invariant subspaces in $\mathcal{H}^{\otimes 2+1}$, $\dim(\mathcal{H})=2$.
S,A denote the symmetry of states (Symmetric or Antisymmetric) with
respect to exchange of clones (first two spaces in the tensor product). d is
the dimension of an invariant subspace.}
\begin{tabular}{|c|c|c|c|}
\hline
space & d & basis & $S_2$\\
\hline
$\mathcal{M}^1$ & $2$ &$-\frac{1}{\sqrt{2}}|0,1,1\rangle
+ \frac{1}{\sqrt{2}}|1,0,1\rangle$ &$A$ \\
 & & $\frac{1}{\sqrt{2}}|0,1,0\rangle - \frac{1}{\sqrt{2}}|1,0,0\rangle$  & \\
\hline
$\mathcal{M}^2$ & $2$ &$\sqrt{\frac{2}{3}}|0,0,0\rangle
+ \sqrt{\frac{1}{6}}|0,1,1\rangle
+\sqrt{\frac{1}{6}}|1,0,1\rangle$ &$S $\\
 & & $\sqrt{\frac{2}{3}}|1,1,1\rangle + \sqrt{\frac{1}{6}}|0,1,0\rangle
 +\sqrt{\frac{1}{6}}|1,0,0\rangle$ & \\
\hline
$\mathcal{M}^3$ &$4$& $|0,0,1\rangle$&$S$\\
 & &$-\frac{1}{\sqrt{3}}|0,0,0\rangle + \frac{1}{\sqrt{3}}|0,1,1\rangle
 + \frac{1}{\sqrt{3}}|1,0,1\rangle$ &\\
 & &$ -\frac{1}{\sqrt{3}}|0,1,0\rangle - \frac{1}{\sqrt{3}}|1,0,0\rangle
 + \frac{1}{\sqrt{3}}|1,1,1\rangle$ &\\
 & & $|1,1,0 \rangle$ &\\
\hline
\end{tabular}
\end{table}

According to what was said before, in this case the operator $P_\mathcal{E}$,
which is now a $8\times 8$ matrix, must be diagonal, as there are no equivalent
representations acting in spaces with the same symmetry of states, i.e.
\begin{equation}
P_\mathcal{E} = a \mathbb{I}_1^1+b \mathbb{I}_2^2+c \mathbb{I}_3^3,
\end{equation}
where $a,b,c \geq 0$.

The known transformation for the optimal cloning of a qubit \cite{buzek},
gives $P_\mathcal{E}$ of the above form with $a=0$, $b=1$, and $c=0$. (One
could also use above formalism to find such a transformation, if it was not
known, by maximizing fidelity with respect to $a$, $b$, and $c$ with
constraints imposed by the trace preserving condition (\ref{eq:trace})). In other
words the operator $P_\mathcal{E}$ has two non-zero eigenvalues both equal
$1$, and the corresponding eigenvectors belong to the two dimensional invariant
subspace $\mathcal{M}^2$, symmetric with respect to permutations of the two
clones.

\subsection{d=3}
In the case $d=3$, the tensor product (\ref{eq:tensor}) decomposes in the
following manner:
\begin{equation}
R_3 \otimes R_3 \otimes R_3^* = R_1 \oplus R_3 \oplus R_3 \oplus R_3 \oplus R_5
\oplus R_5 \oplus R_7.
\end{equation}
We have seven invariant subspaces $\mathcal{H}^{\otimes
2+1}=\oplus_{i=1}^7\mathcal{M}^i$ described in Table \ref{ta:d3}.
\begin{table}[htb]
\caption{\label{ta:d3}Invariant subspaces in $\mathcal{H}^{\otimes 2+1}$, $\dim(\mathcal{H})=3$. 
S,A denote the symmetry of states (Symmetric or Antisymmetric) with
respect to exchange of clones. d is the dimension of an invariant subspace.}
\begin{tabular}{|c|c|c|c|}
\hline
space & d &  basis & $S_2$\\
\hline
$\mathcal{M}^1$ & $1$ &$\frac{1}{\sqrt{6}}(-|0,1,0\rangle - |0,2,1\rangle+|1,0,0\rangle
-|1,2,2\rangle +|2,0,1\rangle +|2,1,2\rangle)$&$A$ \\
\hline
$\mathcal{M}^2$ & $3$ &$\frac{1}{2}(-|0,1,1\rangle  -|0,2,2\rangle +|1,0,1\rangle
+|2,0,2\rangle)$ &$A$ \\
 & & $\frac{1}{2}(|0,1,0\rangle  -|1,0,0\rangle -|1,2,2\rangle +|2,1,2\rangle)$ & \\
 & & $\frac{1}{2}(|0,2,0\rangle  -|1,2,1\rangle -|2,0,0\rangle -|2,1,1\rangle$ & \\
\hline
$\mathcal{M}^3$ &$3$&$\frac{1}{\sqrt{3}}(|0,2,2\rangle - |1,1,2\rangle + |2,0,2\rangle)$ &$S$ \\
 & & $\frac{1}{\sqrt{3}}(-|0,2,1\rangle + |1,1,1\rangle - |2,0,1\rangle)$ &\\
 & & $\frac{1}{\sqrt{3}}(|0,2,0\rangle - |1,1,0\rangle + |2,0,0\rangle)$ &\\
\hline
$\mathcal{M}^4$ &$3$&$\frac{1}{2\sqrt{15}}(6|0,0,0\rangle + 3|0,1,1\rangle + |0,2,2\rangle)
+3|1,0,1\rangle+ 2|1,1,2\rangle+ |2,0,1\rangle$  &$S$ \\
 & &$\frac{1}{2\sqrt{15}}(3|0,1,0\rangle + 2|0,2,1\rangle + 3|1,0,0\rangle)
+4|1,1,1\rangle+ 3|1,2,2\rangle+ 2|2,0,1\rangle+3|2,1,2\rangle$ & \\
 & &$\frac{1}{2\sqrt{15}}(|0,2,0\rangle + 2|1,1,0\rangle + 3|1,2,1\rangle)
+|2,0,0\rangle+ 3|2,1,1\rangle +6|2,2,2\rangle$  & \\
\hline
$\mathcal{M}^5$ & $5$ &$\frac{1}{\sqrt{2}}(-|0,1,2\rangle +|1,0,2\rangle)$ &$A$ \\
 & & $\frac{1}{2}(|0,1,1\rangle  -|0,2,2\rangle -|1,0,1\rangle +|2,0,2\rangle)$ & \\
 & & $\frac{1}{2\sqrt{3}}(-|0,1,0\rangle +2|0,2,1\rangle +|1,0,1\rangle
 -|1,2,2\rangle-2|2,0,1\rangle+|2,1,2\rangle)$ & \\
 & & $\frac{1}{2}(-|0,2,0\rangle  +|1,2,1\rangle +|2,0,0\rangle -|2,1,1\rangle)$ & \\
 & & $\frac{1}{\sqrt{2}}(-|1,2,0\rangle  +|2,1,0\rangle)$ & \\
\hline
$\mathcal{M}^6$ & $5$ &$\frac{1}{\sqrt{6}}(2|0,0,1\rangle +|0,1,2\rangle+|1,0,2\rangle)$ &$S$ \\
 & & $\frac{1}{2\sqrt{3}}(-2|0,0,0\rangle +|0,1,1\rangle +|0,2,2\rangle
 +|1,0,1\rangle+2|1,1,2\rangle+|2,0,2\rangle)$ & \\
 & & $\frac{1}{2}(-|0,1,0\rangle -|1,0,0\rangle +|1,2,2\rangle +|2,1,2\rangle)$   & \\
 & & $\frac{1}{2\sqrt{3}}(-|0,2,0\rangle -2|1,1,0\rangle -|1,2,1\rangle
 -|2,0,0\rangle-|2,1,1\rangle+2|2,2,2\rangle)$ & \\
 & &$\frac{1}{\sqrt{6}}(-2|2,2,1\rangle -|1,2,0\rangle-|2,1,0\rangle)$ & \\
\hline
$\mathcal{M}^7$ & $7$ &$|0,0,2\rangle$ &$S$ \\
 & & $\frac{1}{\sqrt{3}}(-|0,0,1\rangle +|0,1,2\rangle +|1,0,2\rangle)$ & \\
 & & $\frac{1}{\sqrt{15}}(-|0,0,0\rangle -2|0,1,1\rangle +|0,2,2\rangle
 -2|1,0,1\rangle+2|1,1,2\rangle +|2,0,2\rangle )$&\\
 & & $\frac{1}{\sqrt{10}}(|0,1,0\rangle -|0,2,1\rangle +|1,0,0\rangle
 -2|1,1,1\rangle+|1,2,2\rangle-|2,0,1\rangle)+|2,1,2\rangle)$ & \\
 & & $\frac{1}{\sqrt{15}}(|0,2,0\rangle +2|1,1,0\rangle
 -2|1,2,1\rangle +|2,0,0\rangle-2|2,1,1\rangle +|2,2,2\rangle )$&\\
 & & $\frac{1}{\sqrt{3}}(|1,2,0\rangle +|2,1,0\rangle -|2,2,1\rangle)$ & \\
 & & $|2,2,0\rangle$ &\\
\hline
\end{tabular}
\end{table}
One can see that in case of $d=3$ the form of $P_\mathcal{E}$ can be more
complicated. It needs not to be diagonal as we have two equivalent
representations in the subspaces $\mathcal{M}^3$ and $\mathcal{M}^4$, which
are both symmetric under exchange of the clones. The most general $P_\mathcal{E}$
satisfying the covariance condition will consist of one block for
$\mathcal{M}^3$ and $\mathcal{M}^4$ subspaces and will be diagonal in the
remaining part of the space.

Using this picture we can now look at the coherent and the universal cloning
transformations for $d=3$ analyzed earlier. Interestingly, the $27\times 27$
matrices $P_\mathcal{E}$ for both universal and coherent cloning have only
three non-zero eigenvalues, all equal to $1$. The difference between the two
cloning transformations is in the eigenvectors of the corresponding
$P_\mathcal{E}$ operators. In both cases, however, all eigenvectors are
confined to the subspace $\mathcal{M}_3\oplus\mathcal{M}_4$. It means that
the only non-zero elements of $P_\mathcal{E}$ are in the block part considered
above.

Let us denote by $|\phi^i_n\rangle$ the $n$-th basis state in $\mathcal{M}^i$,
as ordered in Table \ref{ta:d3}. In both cases of the universal and the coherent
cloning the three eigenvectors corresponding to the eigenvalue $1$ are given by:
\begin{equation}
|v_k\rangle = a|\phi^3_k \rangle+ b|\phi^4_k\rangle, \quad
k=1,2,3,
\end{equation}
with the coefficients :
\begin{equation*}
\begin{array}{c|c|c}
& $a$ & $b$\\
\hline
\textrm{universal} & \sqrt{\frac{1}{6}} & \sqrt{\frac{5}{6}}\\
\hline
\textrm{coherent} & \sqrt{\frac{1}{2}-\frac{13}{6\sqrt{21}}}&\sqrt{\frac{1}{2}
+\frac{13}{6\sqrt{21}}}
\end{array}
\end{equation*}
Looking at these examples one may try to formulate some conjectures concerning
the operator $P_\mathcal{E}$ for higher dimensions $d$. It seems that it
should have $d$ eigenvalues equal to $1$ and all other $d^3-d$ of them
vanishing. Moreover, eigenvectors corresponding to the non-zero eigenvalues
should be spanned by vectors from $d$-dimensional invariant subspaces,
symmetric with respect to exchange of clones.

\section{Summary}
We have analyzed the problem of cloning spin coherent states using the method
proposed by Gisin and Massar \cite{gisin}. We found an explicit optimal
cloning transformation for spin coherent states in the case when dimension of a
Hilbert space is $d=3$. Maximal attainable fidelities for higher dimensions ($d
\leq 16$) were found. These results allowed us to make some comments regarding
the problem of cloning harmonic oscillator coherent states. In particular our
numerical results for fidelities show that that maximal attainable fidelity for cloning of harmonic
oscillator coherent states is in the range $2/3 \leq \mathcal{F} \leq 0.699$. Additionally we analyzed
more closely the cloning transformation for $d=2$ and $d=3$ with the help of
Jamio{\l}kowski isomorphism between CP maps and positive operators
\cite{aj:positive}. Doing so we followed the general approach towards
nonuniversal covariant cloning, proposed by D'Ariano and Presti
\cite{ga:nonuniversal}. Finally we formulated some conjectures concerning the
structure of CP maps corresponding to optimal spin coherent cloning
transformations in arbitrary dimension d.

\begin{acknowledgments}
K.W. research was partially supported by a KBN Grant 2P03 B 02123 and the
European Commission through the Research Training Network QUEST. M.K. research
was partially supported by VW grant ''Entanglement measures and the influence
of noise''
\end{acknowledgments}

\end{document}